\documentclass[12pt]{article}
\newlength{\vshift}
\input epsf.tex
\newlength{\hshift}

\usepackage{amssymb}
\usepackage{color}
\usepackage{amsmath}
\usepackage{amsthm}
\usepackage{amsopn}
\def\uno{\mbox{1 \kern-.59em {\rm l}}}

\def\p{\partial}

\def\th{\theta}
\def\nn{\nonumber}
\def\be{\begin{equation}}
\def\ee{\end{equation}}
\def\bea{\begin{eqnarray}}
\def\eea{\end{eqnarray}}

\begin{document}

 \vspace*{0cm}

 \begin{center}

  {\bf{\large Neutrino-electron scattering in noncommutative space}}

\vskip 4em

 {$^a${\bf M. M. Ettefaghi} \footnote{ mettefaghi@qom.ac.ir  } and $^b${\bf T. Shakouri}}
 \vskip 1em
$^a$ Department of Physics, The University of Qom, Qom 371614-6611, Iran.

$^b$ Department of physics, Tafresh University, Tafresh, Iran.
 \end{center}

 \vspace*{1.9cm}

\begin{abstract}
Neutral particles can couple with the $U(1)$ gauge field in the
adjoint representation at the tree level if the space-time
coordinates are noncommutative (NC). Considering neutrino-photon
coupling in the NC QED framework, we obtain the differential cross
section of neutrino-electron scattering. Similar to the magnetic
moment effect, one of the NC terms is proportional to $\frac 1 T$,
where $T$ is the electron recoil energy. Therefore, this scattering
provides a chance to achieve a stringent bound on the NC scale in
low energy by improving the sensitivity to the smaller electron
recoil energy.

\end{abstract}
\newpage
\section{Introduction}
There exist strong evidences such as solar and atmospheric as well
as long baseline accelerator and reactor neutrino measurements which
imply finite neutrino masses and mixings \cite{mass}. The finite
neutrino masses leads to the couplings of neutrinos with photons
through loop corrections in the usual space-time. These properties
of neutrinos can be explored using a number of possible physical
processes involving a neutrino with a magnetic moment. Among these
are the neutrino-electron scattering, spin-flavor precession in an
external magnetic field, plasmon decay and the neutrino decay. For
the first process, the magnetic moment contribution of neutrinos in
the differential cross section of the neutrino-electron scattering
is \cite{MM}
\be\label{MM}\frac{d\sigma_{MM}}{dT}=\frac{\pi\alpha^2\mu_\nu^2}{m_e^2}[\frac{1}{T}-\frac{1}{E_\nu}],
\ee where $T$ is electron recoil energy and $\mu_\nu$ is neutrino
magnetic moment which is expressed in unit of $\mu_B$. Clearly this
contribution is dominant at the small recoil energies, {\it i.e.},
the lower the smallest measurable recoil energy is, the smaller
values of the magnetic moment can be probed. To perform such an
experiment either solar or reactor neutrinos have been used. The
MUNU \cite{MUNU} experiment at the Bugey reactor in France and
TEXONO \cite{TEXONO} at the Kuo-Sheng reactor in Taiwan have
analyzed the recoil electron energy spectrum dN/dT for very small
recoil kinetic energies, $T\lesssim 1 MeV$. The limit on the
neutrino magnetic moments of
$\mu_{\bar{\nu}_e}<7.4\times10^{-11}\mu_B$ at 90\% confidence level
was derived \cite{wong}. Moreover, using neutrino-electron
scattering the experimental constraints on non-standard neutrino
interactions and unparticle physics were explored recently by TEXONO
collaboration \cite{TEXONO2}. In this paper we show that this
experiment is appropriate to obtain stringent bound on the
noncommutative scale in low energy.

Noncommutative (NC) quantum field theories have been considered in
the recent decade extensively because of some motivations coming
from string theory \cite{douglas} and quantum gravity
\cite{minimallength}.  In the NC field theory one encounters new
properties such as UV/IR mixing \cite{UVIR}, Lorentz violation
\cite{clock} and CP-violation \cite{charge}. The phenomenological
aspects of the NC field theory at testable energy scales have been
studied extensively:
\begin{itemize}
\item At the atomic scale, for instance, the Lamb shift in the hydrogen atom \cite{lamb}, the positronium hyperfine
splitting \cite{h1} and the transitions in the Helium atom \cite{h2}
were studied in the NC space-time. The best bound on the NC scale
obtained at this scale is about $30 GeV$ from the transitions in the
Helium atom.
\item At the electroweak scale, for instance, $Z\rightarrow\gamma\gamma$ \cite{Ztogammagamma}, $Z\rightarrow l^+l^-$
and $W\rightarrow \nu_l l$ \cite{SMscale1}, quarkonia decay
\cite{SMscale2}, top quark decay \cite{SMscale3} and so on were
studied in the NC standard model framework. In this scale an
experimental bound on the NC scale about $141 GeV$ was found by the
OPAL collaboration using $e^+e^-\rightarrow \gamma\gamma$ at LEP
\cite{OPAL}.
\item At the above the electroweak scale accessible in future
experiment, the NC effects were explored for various processes such
as $e^+e^-$ scattering \cite{ee}, hadrons colliders \cite{LHC} and
photon-photon colliders \cite{photon-photon}.
\end{itemize}
Also the NC signatures were followed in the astrophysics and
cosmology, for example see \cite{cosmology}. The usual bounds on the
NC scale obtained from the mentioned considerations are about $1
TeV$. However, there exist various candidates of new physics such as
supersymmetry at this scale and searching for NC signals at the
energies near this scale seems ambiguous. Therefore, the study of
the NC signals which are dominant below the electroweak scale will
be important. The NC field theories are constructed on the
space-time coordinates which are operators and do not obey
commutative algebra. In the case of canonical version of the NC
space-time, the coordinates satisfy the following algebra: \be
\th^{\mu\nu}=-i[\hat{x}^\mu,\hat{x}^\nu], \ee where a hat indicates
a NC coordinate and $\th^{\mu\nu}$ is a real, constant and
antisymmetric matrix.
 To construct the NC field theory, according to the Weyl-Moyal correspondence, an ordinary function can
  be used instead of the corresponding NC one by replacing the ordinary product with the star product as follows:
\be\label{starproduct}
 f\star
g(x,\theta)=f(x,\theta)\exp(\frac{i}{2}\overleftarrow{\partial}_\mu
\theta^{\mu\nu}\overrightarrow {\partial}_\nu)g(x,\theta). \ee Due
to the above correspondence a neutral particle (as well as a charged
particle) can couple with the $U(1)$ gauge field in the adjoint
representation. Some effects of this new coupling were studied in
the literature \cite{neutral boson,pho-neu,hez,massive neutrino}. In
particular, in \cite{pho-neu} the coupling between photons and
left-(right-)handed neutrinos was considered. The usual requirement
that any new energy-loss mechanism in globular stellar clusters
should not excessively exceed the standard neutrino losses implies a
scale of NC gauge theory above the scale of week interactions. In
this paper, we study the effects of this new coupling of neutrinos
on the neutrino-electron scattering. As we will see, the behavior of
the NC corrections in terms of $T$ (electron recoil energy) is
similar to the correction due to the neutrino magnetic moment which
is dominant for the small electron recoil energy. Therefore, we
compare the NC contribution in the neutrino-electron scattering with
the standard model one for reactor neutrino whose energies are about
$1 MeV$ and for neutrinos whose energies are about $1 GeV$ such as
beta beam \cite{betabeam}. This paper is organized as follows: In
Section two we review briefly the NC QED for neutral particles. In
Section three the neutrino-electron scattering is discussed.
Finally, we summarize our results in the last section.

\section{Neutral particles in the NC QED}
According to the Weyl-Moyal correspondence, Eq. (2), the usual
products in $eA_\mu\psi$ where $e$, $A_\mu$, and $\psi$ are the
coupling, the gauge field, and the matter field, respectively, are
replaced by star products. This leads to an ambiguity in the
ordering of fields: $eA_\mu\star\psi$, $e\psi \star A_\mu$, and
$e(A_\mu\star\psi-\psi \star A_\mu)$. In the action, however, it has
been shown that the two first couplings are the charge conjugations
of each other, but the third one is the charge conjugation of itself
\cite{charge}. Therefore, a neutral particle can have the third
coupling and its covariant derivative is defined as follows: \be
\hat{D}_\mu\hat{\psi}=\p_\mu\hat{\psi}-ie(\hat{A}_\mu\star\hat{\psi}-\hat{\psi}\star
\hat{A}_\mu),\ee where hats on the fields are used to emphasize that
these fields are defined in the NC space-time. We can write these NC
fields in terms of the usual fields using corresponding
Seiberg-Witten maps. The corresponding Seiberg-Witten  maps up to
the first order of $\th$ are \cite{swmap} \bea
\hat{\psi}=&&\!\!\!\!\!\!\!\psi+\th^{\mu\nu}A_\nu\p_\mu\psi,\nn\\
\hat{A}_\mu=&&\!\!\!\!\!\!\!A_\mu+e\th^{\nu\rho}A_\rho[\p_\nu
A_\mu-\frac 1 2 \p_\mu A_\rho].\eea Hence, the action describing a
neutral fermion field in the NC QED framework is \be S=\int
d^4x(\bar{\hat{\psi}}\star
i\gamma^\mu\hat{D}_\mu\hat{\psi}-m\bar{\hat{\psi}}\star\hat{\psi}).\ee
After using above Seiberg-Witten maps and expanding the star product
up to the first order of $\th$, this action can be written as
follows \cite{pho-neu}: \be S=\int
d^4x\bar{\psi}[(i\gamma^\mu\p_\mu-m)-\frac e 2
\th^{\nu\rho}(i\gamma^\mu(F_{\nu\rho}\p_\mu+F_{\mu\nu}\p_\rho+F_{\rho\mu}\p_\nu)-mF_{\nu\rho})]\psi,\ee
where $F_{\mu\nu}=\p_\mu A_\nu-\p_\nu A_\mu$. The NC induced
photon-neutrino vertex in the case of massless left-handed neutrino
is \be\label{rule}
\Gamma^\mu(\nu\bar{\nu}\gamma)=-e\th^{\mu\nu\rho}k_\nu
q_\rho(\frac{1-\gamma_5}{2}),\ee where
$\th^{\mu\nu\rho}=\th^{\mu\nu}\gamma^\rho+\th^{\nu\rho}\gamma^\mu+\th^{\rho\mu}\gamma^\nu$.
Since the expansion of the action in terms of $\th$ is truncated up
the first order of $\th$, it is permissible to apply this theory for
the energies below NC scale.

The Weyl-Moyal correspondence leads to some restrictions on NC gauge
theories termed as NC gauge theory no-go theorem. These restrictions
cause one to have some problems for constructing the NC standard
model \cite{no-go}. Until now, two approaches have been suggested to
solve these problems. In one of them, the gauge group is restricted
to $U(n)$ and the symmetry group of the standard model is achieved
by the reduction of $U(3)\times U(2)\times U(1)$ to $SU(3)\times
SU(2)\times U(1)$ by an appropriate symmetry breaking \cite{nNCSM}.
In the other approach, the $SU(n)$ gauge group in the NC space-time
can be achieved via Seiberg-Witten map \cite{swmap}. Hence, the NC
standard model is constructed through replacing the usual products
and fields, respectively, by star products and NC fields which can
be written in terms of the usual fields using the corresponding
Seiberg-Witten maps \cite{sm}. The NC standard model based on the
$U(3)\times U(2)\times U(1)$ gauge group incorporates directly a
coupling between photon and left-handed neutrinos. But the coupling
between the left-handed neutrinos and photon cannot be accommodated
in the NC standard model based on the $SU(3)\times SU(2)\times U(1)$
gauge group since the left-handed neutrinos are involved in the
$SU(2)$ gauge theory \cite{hez}. The right-handed neutrinos which
are singlet under the standard model gauge transformation can be
added and couple directly with photon in both version of the NC
standard model.

\section{Neutrino-electron scattering}
We are interested in the first order of NC corrections on the cross
section of the neutrino-electron scattering. In the usual
space-time, $\nu_\mu+e\rightarrow\nu_\mu+e$, where $\nu_\mu$ stands
for either muon neutrino or antimuon neutrino, proceeds solely via
neutral current channel (via $Z_0$ exchange) while
$\nu_e+e\rightarrow\nu_e+e$ proceeds via both neutral and charged
current channels (via both $Z_0$ and $W^{\pm}$ exchange). Hence, the
corresponding cross sections are proportional to $G_F^2$. In the NC
standard model,
 the interference of a
standard model diagram with a diagram where one electroweak vertex
is replaced by the first order in $\th$ is zero. That is why there
is $\frac{\pi}{2}$ phase difference between the usual terms and the
first order of $\th$ NC terms \cite{sm} and it causes ${\cal
M}_1^\ast {\cal M}_2+{\cal M}_1{\cal M}^\ast_2$ to be zero in the
case of massless neutrinos. Therefore, the leading order term of the
NC standard model corrections on the neutrino-electron scattering is
proportional to $G^2_F\th^2$. However, when we include the NC
induced photon-neutrino coupling, there exists a new channel at the
tree level which proceeds via photon exchange. The interference
between this new channel and electroweak channels is also zero and
the first order of the NC corrections due to the photon exchange
channel is proportional to $\th ^2$. In fact the leading order term
of the NC correction for neutrino-electron scattering is
proportional to $\th^2$ in both NC QED and NC standard model
framework. However, the latter is suppressed by a factor $G_F^2$ in
comparison with the former.

In the standard model, the differential cross sections for all
neutrinos-electron scattering are given by
\cite{n-elitrature}\be\label{smcr}
\frac{d\sigma_l}{dT}=\frac{2G_F^2m_e}{\pi
E_\nu^2}(a^2E_\nu^2+b^2(E_\nu-T)^2-abm_eT),\ee where subscript $l$
refers to neutrino flavors and $E_\nu$ and $T$ stand for the energy
of the incident neutrino and the kinetic energy of the recoil
electron, respectively. $a$ and $b$ are process-dependent parameters
and are given by table (1) as functions of the weak mixing angel,
$\th_W$.
\begin{small}
\begin{table}
\begin{center}
\begin{tabular}{|c|c|c|c|c|}
                             \hline
                              & $\nu_e+e\rightarrow\nu_e+e$ & $\bar{\nu}_e+e\rightarrow\bar{\nu}_e+e$
                              &$\nu_l+e\rightarrow\nu_l+e$  & $\bar{\nu}_l+e\rightarrow\bar{\nu}_l+e$ \\
                             \hline
                             $a$ &$-\frac{1}{2}-s^2$ & $-s^2$ & $\frac 1 2 -s^2$ & $-s^2$ \\
                             \hline
                             $b$ & $-s^2$ & $-\frac 1 2 -s^2$ & $-s^2$ & $\frac 1 2 -s^2$ \\
                             \hline
                           \end{tabular}
\end{center}
\caption{Standard model $a$ and $b$ parameter values for the
differential cross-section, given by Eq. (\ref{smcr}). Here
$s=\sin\th_W$ where $\th_W$ is the electroweak mixing angle.}
\end{table}
\end{small}

\begin{figure}
\centerline{\epsfysize=2.5in\epsfxsize=3in\epsffile{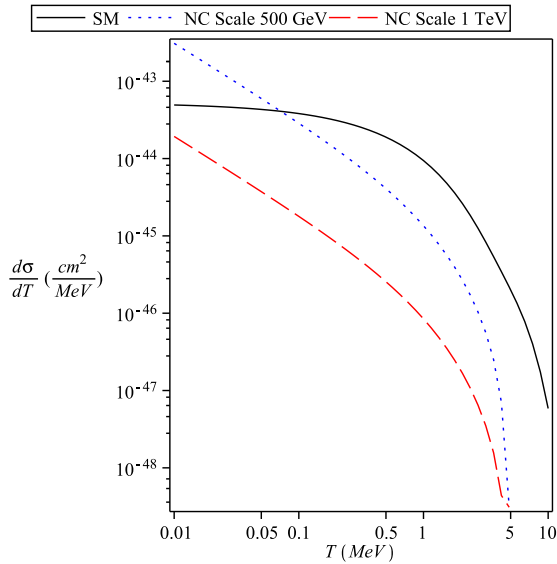}}\caption{The
differential neutrino-electron cross section versus the electron
recoil energy in $E_\nu$ of the order of a few $MeV$ (reactor
neutrino). Black (solid), blue (dot), and red (dashed) curves
represent the contributions of the standard model, the NC QED with
$\Lambda_{NC}=500 GeV$, and the NC QED with $\Lambda_{NC}=1 TeV$,
respectively.
 }
\label{fig1}
\end{figure}

Using Eq. (\ref{rule}), we can write the Feynman amplitude of the NC
QED contribution to the neutrino-electron scattering as follows: \be
-i{\cal M}_{NC}=\frac{e^2}{2q^2}[\bar{u}(p^\prime)\gamma^\mu
u(p)][\bar{u}(k^\prime)\th_{\mu\nu\rho}k^\nu q^\rho
(\frac{1-\gamma_5}{2})u(k)].\ee Unitarity is satisfied for
$\th_{0i}=0$ and $\th_{ij}\neq0$ \cite{unitarity}. Hence, let us
assume $\th_{i0}=0$ and define
$\vec{\th}=(\th_{23},\th_{31},\th_{12})$. Also we ignore neutrino
mass. Summing over initial and averaging over final spin stats, one
can obtain \be |{\cal
M}_{NC}|^2=\frac{32e^4}{q^4}(\frac{\vec{\th}.(\vec{k}\times
\vec{k}^\prime)}{2})^2\{(p.k)(p^\prime.k^\prime)
+(p.k^\prime)(p^\prime.k)-m^2_e(k.k^\prime)\},\ee in which
$\th_{\mu\nu}k^\mu{k^\prime}^\nu=\vec{\th}.(\vec{k}\times\vec{k^\prime})$
and Dirac equation for neutrinos, $k\!\!\!/u(k)=0$ and
$\bar{u}(k^\prime)k^\prime\!\!\!\!\!/=0$, are used. We choose the
following orientations for the lab frame: \bea p&=&(m_e,0,0,0),\nn\\
k&=&(E_\nu,0,0,E_\nu),\nn\\
p^\prime&=&(T+m_e,|\vec{p^\prime}|\sin\alpha\cos\phi,|\vec{p^\prime}|\sin\alpha\sin\phi,|\vec{p^\prime}|\cos\alpha),\nn\\
k^\prime&=&(E_\nu-T,-|\vec{p^\prime}|\sin\alpha\cos\phi,-|\vec{p^\prime}|\sin\alpha\sin\phi,E_\nu-|\vec{p^\prime}|\cos\alpha),\nn\\
\vec{\th}&=&(\th\sin\lambda,0,\th\cos\lambda).\eea After averaging
over $\lambda$, the contribution of the NC QED to the differential
cross section of the neutrino-electron scattering is obtained as
follows:
 \be\label{NC}
\frac{d\sigma_{NC}}{dT}=\frac{e^4E_\nu^2\th^2}{16\pi}[\frac{1}{T}-\frac{2}{E_\nu}+
\frac{3T-2m_e}{2E_\nu^2}-\frac{T^2-2m_eT}{2E_\nu^3}-\frac{m_eT^2}{4E_\nu^4}(1-\frac{m_e}{E_\nu})].\ee
The first term of this equation is dominant at low recoil energy of
electron, similar to the contribution of the neutrino magnetic
moment, see Eq. (\ref{MM}). Explicitly, the NC QED contribution will
exceed the standard model one for recoil energies \be \frac T
m_e<\frac{\pi^2\alpha^2E_\nu^2\th^2}{G_F^2m_e^2}=\frac{\pi^2\alpha^2E_\nu^2}{G_F^2m_e^2\Lambda_{NC}^4},
\ee where $\Lambda_{NC}=\frac{1}{\sqrt{\th}}$. For example, let us
consider electron antineutrinos with energies up to about $10 MeV$
which are emitted by nuclear reactors. The exact energy spectrum
depends on the specific fuel composition of the reactor. For
instance, the energy spectrum of neutrinos coming from the
fissioning of $^{235}U$ is approximately given by \cite{reactor} \be
\frac{dN_\nu}{dE_\nu} \sim \exp (0.870 - 0.160E_\nu -
0.0910E^2_\nu), \ee where the antineutrino energy, $E_\nu$, is given
in $MeV$. Now, the relevant quantity for the number of
neutrino-electron elastic scattering events within the interval $[T
,T + dT]$ is the cross section folded with the above energy spectrum
and is given by \be
\langle\frac{d\sigma}{dT}\rangle=\int^\infty_{E_\nu^{min}}\frac{dN_\nu}{dE_\nu}\frac{d\sigma}{dT}dE_\nu,\ee
where ${E_\nu^{min}}=0.5(T+\sqrt{T^2+2Tm_e})$. The behavior of
$\langle\frac{d\sigma_{\bar{\nu}_e}}{dT}\rangle$ and
$\langle\frac{d\sigma_{NC}}{dT}\rangle$ for $\Lambda_{NC}$ equals
$500 GeV$ and $1 TeV$ versus $T$ in the range $0.01MeV\leq T\leq10
MeV$ is depicted in Figure \ref{fig1}.
 For the NC scale about $500
GeV$, the NC contribution is more than the commutative one for
$T<0.1 MeV$. Therefore, using an analysis similar to TEXONO
\cite{wong}, one can obtain a bound of a few hundred of $GeV$ on the
NC scale. It is noticeable that this bound on the NC scale is
obtained by an experiment at energy of a few of $MeV$. However one
can obtain more stringent bounds if the sensitivity to the smaller
electron recoil energy is improved.

\begin{figure}
\centerline{\epsfysize=2.5in\epsfxsize=3in\epsffile{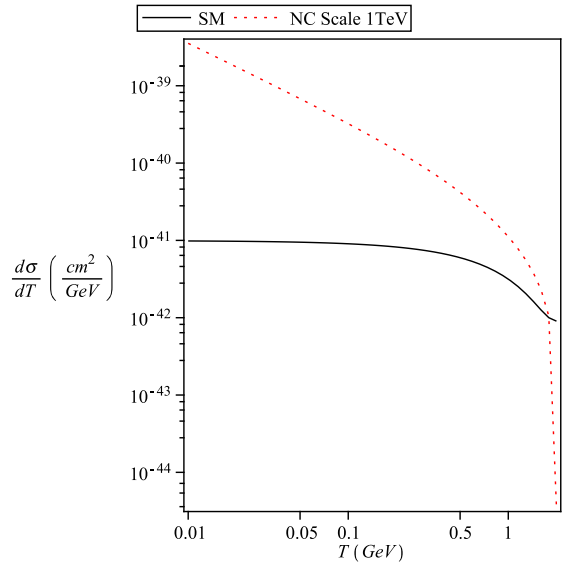}}\caption{The
differential neutrino-electron cross section versus the electron
recoil energy in $E_\nu$ of the order of a few $GeV$. Black (solid),
and red (dot) curves represent the contributions of the standard
model, and the NC QED with $\Lambda_{NC}=1 TeV$, respectively.
 }
\label{fig2}
\end{figure}

Moreover, there is a difference between the contribution of the NC
QED and the contribution of the magnetic moment to the cross section
of the neutrino-electron scattering; $\frac{d\sigma_{NC}}{dT}$ is
proportional to $E_\nu^2$ and grows rapidly with energy\footnote{It
seems to conflict with unitarity theorem. However, we should remind
that the corresponding Lagrangian has been expanded in terms of NC
parameter, $\th$. Actually, the region where the $\th$-expansion is
well defined is restricted to $\th^{\mu\nu}p^\mu q^\nu <1$ or
$\frac{E}{\Lambda_{NC}}<1$ where $E=\sqrt{s}$ is the energy of the
system.}. Hence, the NC contribution can be more significant in
higher energies. For instance, we contrast
$\frac{d\sigma_{\bar{\nu}_e}}{dT}$ and $\frac{d\sigma_{NC}}{dT}$ at
$E_\nu=2 GeV$ and $\Lambda_{NC}=1 TeV$ in Figure \ref{fig2}.
Therefore, using neutrino beams with energy about $1 GeV$ such as
beta beam neutrinos \cite{betabeam}, one can obtain a bound on
$\Lambda_{NC}$ more stringent than $1 TeV$ which is achievable by
future linear accelerators such as LHC.

\section{conclusion}
Neutrino-electron scattering is one of the various processes which
can be used to study the neutrino electromagnetic properties. In the
NC field theory neutrinos, as well as the others neutral and charged
particles, can have new electromagnetic interactions. Namely,
neutrinos can couple with photon through adjoint representation in
the NC QED. In this paper we have calculated the NC QED corrections
on the neutrino-electron scattering. This scattering is exceptional
from the others NC phenomenologies in the sense that the behavior of
the NC QED contributions in terms of electron recoil energy is
similar to that of the neutrino magnetic moments in the commutative
space-time, {\it i.e.}, one of NC correction terms is proportional
to the inverse of the electron recoil energy. In contrast to the
linear accelerators in which we need higher energy to find more
stringent bound on the NC scale, for neutrino-electron scattering
the crucial quantity is the minimum electron recoil energy
accessible to the experiment. With current Experiments such as
TEXONO \cite{TEXONO}, in which reactor neutrinos with energy about
$1 MeV$ are used, one can obtain a bound about a few hundred of
$GeV$.

\vspace{5mm} {\bf Acknowledgement:}
 Authors would like to thank M. Hghaighat and
R. Moazemi for reading the manuscript and their useful suggestions.
The financial support of the University of Qom research council is
acknowledged.

\end{document}